\documentclass[aps,10pt,twocolumn,showpacs]{revtex4}
\usepackage{amsfonts}
\usepackage{amsmath}
\usepackage{amssymb}
\usepackage{verbatim}
\usepackage[dvips]{graphicx}

\begin{document}

\title{Ground-state magneto-optical resonances in Cesium vapour
confined in an extremely thin cell }
\author{
C. Andreeva$^{1}$}
 \author{A. Atvars$^{2}$}
 \author{M. Auzinsh$^{2}$}
 \author{K.Bluss$^{2}$}
  \email{stefka-c@ie.bas.bg}\author{ S. Cartaleva$^{1}$}
 \author{L. Petrov$^{1}$}
 \author{D. Slavov$^{1}$}
\affiliation{
 $^{1}$ Institute of Electronics - Bulgarian Academy of  Sciences, boul. Tzarigradsko shosse 72, 1784 Sofia,
 Bulgaria\\
  $^{2}$Department of Physics and
Institute of Atomic Physics and Spectroscopy, University of Latvia,
19 Rainis Blvd.,
LV-1586 Riga, Latvia \\
 }

\begin{abstract}
Experimental and theoretical studies are presented related to the
ground-state magneto-optical resonance prepared in Cesium vapour
confined in an Extremely Thin Cell (ETC, with thickness equal to
the wavelength of the irradiating light). It is shown that the
utilization of the ETC allows one to examine the formation of a
magneto-optical resonance on the individual hyperfine transitions,
thus distinguishing processes resulting in  dark (reduced
absorption) or bright (enhanced absorption) resonance formation.
We report on an experimental evidence of the bright
magneto-optical resonance sign reversal in Cs atoms confined in
the ETC. A theoretical model is proposed based on the optical
Bloch equations that involves the elastic interaction processes of
atoms in the ETC with its walls resulting in depolarization of the
Cs excited state which is polarized by the exciting radiation.
This depolarization leads to the sign reversal of the bright
resonance. Using the proposed model, the magneto-optical resonance
amplitude and width as  a function of laser power are calculated
and compared with the experimental ones. The numerical results are
in good agreement with the experiment.

\end{abstract}

\date{\today }
\maketitle

\section{Introduction}

Recently, the high resolution spectroscopy of alkali atoms confined
in a thin cell has proven very promising for the studies not only of
atom-light, but also of atom-surface interactions \cite{Ham05}. The
realization of such atomic layers has become possible through the
development of the so called Extremely Thin Cells (ETC). The ETC
consists of a thin layer with a thickness of less than 1 $\mu$m and
typical diameters of a few centimeters of gas \cite{Sar01}, which
causes a strong anisotropy in the atom-light interaction.

This anisotropy leads to two main effects in the observed
absorption and fluorescence spectra. First, most of the atoms with
a velocity component normal to the cell walls collide with the
walls before completing the absorption - fluorescence cycle with
the incoming light. Hence, these atoms give a smaller contribution
to the atomic fluorescence compared to atoms flying parallel to
the cell walls. Since generally the light propagation direction is
perpendicular to the cell windows, the Doppler broadening of the
hyperfine (hf) transitions is significantly reduced. Thus, the hf
transitions, which are strongly overlapped in ordinary (cm-size)
cells, in the ETC are well resolved with single-beam spectroscopy.
Unlike in saturation spectroscopy, the sub-Doppler absorption in
an ETC is linear up to relatively large laser power densities,
which allows, for example, the determination of transition
probabilities of the different hf transitions \cite{Sar03}. The
absence of cross-over resonances can also be advantageous when
investigating more complex spectra. Second, the investigation of
atoms confined in cells whose thickness $L$ is comparable to the
wavelength of the light $\lambda$ leads to the observation of
interesting coherent effects. It has been shown that the width of
the hf transitions in absorption changes periodically with cell
width, with minima at $L = (2n + 1) \lambda/2$ \cite{Dut03},
because of the realization of the Dicke regime \cite{Rom55} in the
optical domain. Moreover, for the ETC a significant difference in
the absorption and fluorescence spectra can be observed because
the faster atoms with sufficient interaction time for the
absorption of a photon, but not enough for its subsequent release,
will contribute to the absorption signal but not to the
fluorescence signal \cite{Sar01,Sar04}.

Coherent Population Trapping (CPT) resonances prepared in the Hanle
configuration have been widely investigated for Cs and Rb atoms
confined in ordinary cells. The alkali atoms, situated in the
magnetic field $B$ orthogonal to the atomic orientation/alignment
and scanned around $B$ = 0, are irradiated by monochromatic laser
field in such a configuration that different polarization components
of the light couple the atomic Zeeman sublevels of one of the
ground-state hf levels through a common excited one and introduce
coherence between ground magnetic sublevels at $B$ = 0. As has been
shown in \cite{Dan00,Ren01,Pap02,Aln03,Aln01b,And02}, for degenerate
two-level systems in the absence of depolarizing collisions of the
excited state, Electromagnetically-Induced Transparency (EIT, dark
magneto-optical) resonance or Electromagnetically-Induced Absorption
(EIA, bright magneto-optical) resonance can be observed, depending
on the ratio of the degeneracies of the two states involved in the
optical transition. The EIT resonance is realized when the condition
$F_{g} \rightarrow F_{e} = F_{g} - 1$, $F_{g}$ is met, while the EIA
resonance is observed for $F_{g} \rightarrow F_{e} = F_{g}+ 1$ type
of transitions. Here, $F_{g}$ and $F_{e}$ are the hf quantum numbers
of the ground- and excited-state hf levels, respectively.

The significant disadvantage of magneto-optical resonance studies in
ordinary cells is that because of the Doppler broadening in thermal
cells, a strong overlapping takes place between hf transitions
responsible for dark and bright resonances. Therefore, it is
important to make use of the fact that in an ETC the hyperfine
transitions are well resolved, and the coherent effects can be
investigated under better-defined conditions than in the case of
ordinary cells. Moreover, the  possibility to separate the
contribution of slow and fast atoms to the magneto-optical signal,
provided by using the ETC, will result in a better understanding of
the transient processes in the formation of the coherent resonances.
Besides, magneto-optical resonance investigations in ETCs could give
information on the influence of the cell walls on the atomic
polarization because, also in the magneto-optical signal, the main
contribution is expected from the atoms flying a "long way" along
the cell window.

In this work we present experimental and theoretical results
concerning the ground-state magneto-optical resonances on the
D$_{2}$ line of Cs, obtained at the individual hf transitions
starting from $F_{g}$ = 3 and $F_{g}$ = 4. It is shown that in the
ETC (with thickness equal to the light wavelength), dark (reduced
absorption) magneto-optical resonances are observed at $F_{g}
\rightarrow F_{e} = F_{g} - 1$, $F_{g}$ transitions, which is
similar to the results obtained in the ordinary cells. A very
interesting result is obtained at $F_{g} \rightarrow F_{e} = F_{g} +
1$ transitions. At these transitions, in the dilute Cs vapour
confined in the ETC, bright resonance sign reversal is evidenced,
which is attributed to the depolarization of the Cs atoms' excited
levels. A theoretical model has been developed based on the optical
Bloch equations and involving the elastic interaction processes of
atoms in an ETC with walls that affect the polarization of the
atomic excited levels. The numerical results are in good qualitative
agreement with the experimental ones.

\section{Cesium energy levels and experimental setup}

\begin{figure}[h]
\centering
\includegraphics[height=5cm, width=7cm, scale=0.5]{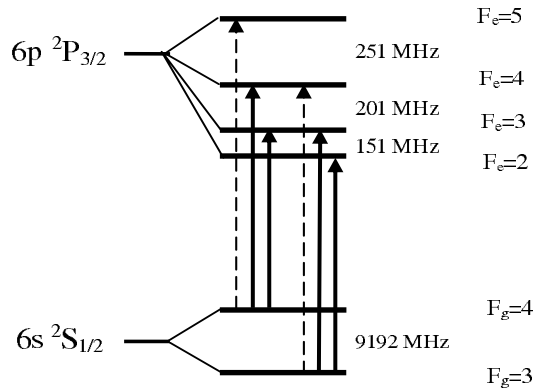}
\caption{ Energy-level diagram for the D$_{2}$ line of $^{133}$Cs.
$F_{g} \rightarrow F_{e} \leq F_{g}$ transitions (solid line) are
distinguished from $F_{g} \rightarrow F_{e} > F_{g}$ transitions
(dashed line). }
 \label{Fig1}
\end{figure}

The relevant Cs energy levels and hf transitions involved are
illustrated in Fig.\ref{Fig1}. The two groups of hf transitions
(starting from $F_{g}$ = 3 and $F_{g}$ = 4) are denoted. The hf
transitions responsible for the formation of magneto-optical
resonances of different sign are distinguished. It should be noted
that the Doppler broadening ($\sim$ 400 MHz) in the ordinary cell,
of the optical hf transitions is larger than the separation
between the excited state hf levels. As a result, in the ordinary
cell the profiles of the three hf transitions starting from a
single ground hf level strongly overlap and form a single
absorption (fluorescence) line. Hence, the three different types
of hf transitions, which are responsible for the formation of both
dark and bright resonances, are involved in a single fluorescence
line. When  exciting atoms in the ordinary cell by mono-mode laser
light and performing the experiment \cite{And02} in dilute Cs
vapor (thus avoiding velocity changing collisions between Cs
atoms), the three hf transitions forming the fluorescence line
will be excited independently, each one at different velocity
class of atoms. Under these conditions the open transitions are
depleted because of the hf optical pumping to the ground hf level
that does not  interact with the laser light. Thus they do not
play a significant role in the formation of the magneto-optical
resonance. At the same time, the closed transition that does not
suffer hf optical pumping mainly determines the sign of the
magneto-optical resonance. Thus, in the ordinary cells dark
resonances are observed at the fluorescence line starting from the
$F_{g}$ = 3 and bright resonances at the fluorescence line
starting from the $F_{g}$ = 4 levels. The dark resonance observed
on the fluorescence line starting from $F_{g}$ = 3 has been
attributed to the $F_{g} = 3 \rightarrow   F_{e}= 2$ closed
transition contribution, while the bright resonance is related to
the $F_{g} = 4 \rightarrow F_{e} = 5$ closed transition.

\begin{figure}[h]
\centering
\includegraphics[height=11cm, width=8.5cm, scale=0.5]{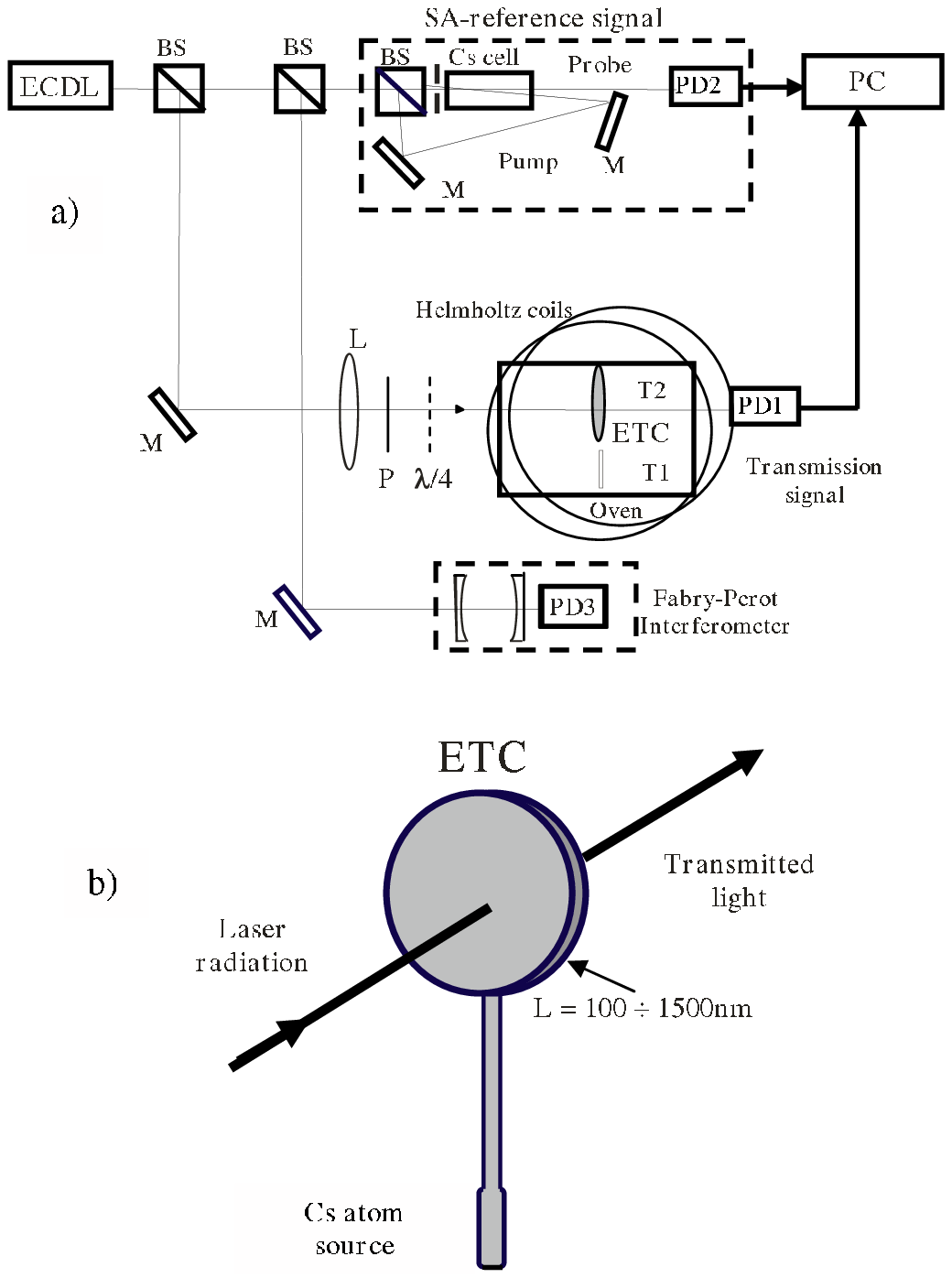}
\caption{ Experimental set up (a) and sketch of ETC (b). }
 \label{Fig2}
\end{figure}

The experimental setup used is schematically presented in
Fig.\ref{Fig2}a. A cw extended cavity diode laser (ECDL) was used
as radiation source. It was operated in single-frequency mode with
$\lambda$  = 852 nm and linewidth of about 3 MHz. Thus, because of
the narrow bandwidth of the laser, it was possible to excite
separately the different hyperfine transitions of the D$_{2}$ line
of $^{133}$Cs atoms [6S$_{1/2}$($F_{g}$ = 3) $\rightarrow$
6P$_{3/2}$($F_{e}$ = 2, 3, 4) or 6S$_{1/2}$($F_{g}$ = 4)
$\rightarrow$ 6P$_{3/2}$($F_{e}$ = 3, 4, 5)]. The main part of the
laser beam was directed at normal incidence onto the ETC (zoomed
in Fig.\ref{Fig2}b) with a source containing Cs. The ETC operates
with a specially constructed oven which maintains a constant
temperature gradient between the Cs atom source (temperature
T$_{1}$) and the cell windows (temperature T$_{2}$), T$_{2}$ $>$
T$_{1}$. The Cs vapour density is controlled by changing the
source temperature. The ETC is placed between a pair of coils in
Helmholtz configuration, which allows application of a scanned
around $B$ = 0 magnetic field in a direction orthogonal to the
laser beam propagation direction. No shielding against laboratory
magnetic field is provided. The light transmitted through the ETC
is measured by photodiode PD1.

Most of the experiments were performed by irradiating Cs atoms by
means of linearly polarized light. In this case the scanned magnetic
field is oriented in a direction orthogonal to the laser beam and to
the light polarization. When light with circular polarization is
used, a quarterwave plate is inserted before the ETC, after the
polarizer P.

First, an experiment was performed aimed at general observation of
the magneto-optical resonance formation along the entire absorption
line profile. In this case, the laser frequency was scanned along
the absorption profile with a rate of about two orders of magnitude
less than that of the magnetic field modulation around $B$ = 0.
During the second part of the experiment related to the
magneto-optical resonance study, the light frequency was fixed at
the center of the examined hf transition and Cs atom absorption was
measured as a function of the magnetic field, which was scanned
around $B$ = 0.

 The remaining part of the laser beam was separated
into two parts and was used for laser frequency control: one part
was sent to a scanning Fabry-Perot interferometer in order to
monitor the single-mode operation of the ECDL and the second one to
an additional branch of the setup including an ordinary, 3 cm long
Cs cell, for simultaneous registration of the Saturated Absorption
(SA) spectrum, which was used for precise scaling and reference of
the laser frequency.

\section{Absorption spectra at the ETC thickness L=$\lambda$/2 and L=$\lambda$}

\begin{figure}[h]
\centering
\includegraphics[height=7.5cm, width=8.5cm, scale=0.5]{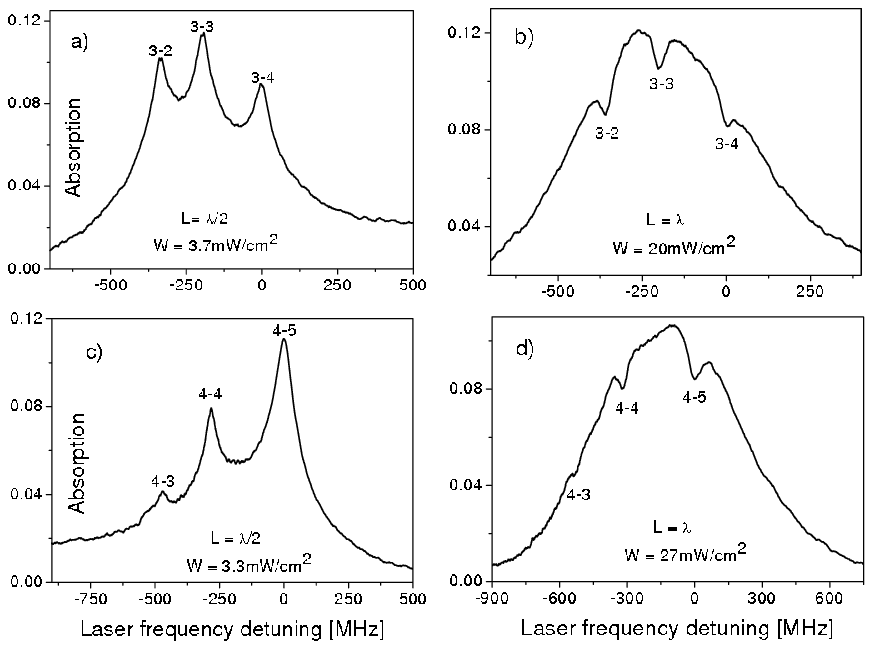}
\caption{Absorption spectra at $F_{g}$=3 (a,b) and $F_{g}$ = 4
(c,d) sets of hf transitions, for ETC thickness $L$=$\lambda$ /2
(a,c) and $L$ = $\lambda$ (b,d).}
 \label{Fig3}
\end{figure}

In order to help the further discussion related to magneto-optical
resonances observed in the ETC, in this Section we present the
absorption spectra of both absorption lines at two thicknesses of
the ETC (Fig.\ref{Fig3}), observed without application of the
magnetic field. For $L$ = $\lambda$/2, the absorption spectra of
the lines starting from $F_{g}$ = 3 and $F_{g}$ = 4 are
illustrated in Fig.\ref{Fig3}a,c. It can be seen that the spectral
profiles of all hf transitions are well resolved due to the
significant enhancement of atomic absorption at the hf transition
center and its reduction toward the wings. As was mentioned in the
introduction, the origin of the narrowing of the hf transition
profiles is attributed to the Dicke effect. The processes
responsible for the coherent Dicke narrowing of the hf transition
profile at $L$ = $\lambda$/2 can be briefly summarized as follows
\cite{Dut03,Sar04,Mau05}. If an atom at the moment of leaving the
cell wall is excited by resonant light, the excitation will start
to precess in phase with the exciting electromagnetic field at the
wall position. However, with the atomic motion the excitation will
drift gradually out of phase with the local exciting field. The
phase mismatch appearing on the line center under a weak exciting
field is independent of the atomic velocity and, for a cell
thickness up to $\lambda/2$, all regions of the cell interfere
constructively, leading to a strong absorption enhancement at the
hf transition center. However, if the exciting light is detuned
from the hf transition center, the angular precession of the
atomic excitation becomes velocity-dependent resulting in a smooth
reduction of the absorption in the wings of the hf profile.

Unlike the atomic spectra observed at L = $\lambda/2$, when the
cell thickness rises up to L = $\lambda$, a completely different
behaviour of the hf transition absorption occurs (Fig.\ref{Fig3}b,
d). Under this condition the Dicke narrowing vanishes and at low
light power density the Dopler profiles of hf transitions forming
the fluorescence line are completely overlapped. However, starting
from a power density $W$ of several mW/cm$^{2}$, narrow dips of
reduced absorption occur, centered at each hf transition. The
observed dips are attributed to velocity-selective saturation of
the hf transition and to velocity-selective population loss due to
optical pumping to the ground-state level that does not interact
with the exciting light field. To complete saturation and/or
optical pumping, the atom needs a longer interaction time with the
exciting light than for a single act of absorption \cite{Bri99}.
Due to this, primarily the very slow atoms or those flying
parallel to the ETC windows can contribute to the reduced
absorption dips at L = $\lambda$. Note that it is expected that
the latter group of atoms will be responsible for the
magneto-optical resonance formation, because this process is also
slower than ordinary linear absorption.

\section{Magneto-optical resonances observation in an ETC with thickness L = $\lambda$:
The Bright resonance sign reversal}

\begin{figure}[h]
\centering
\includegraphics[height=15cm, width=7cm, scale=0.5]{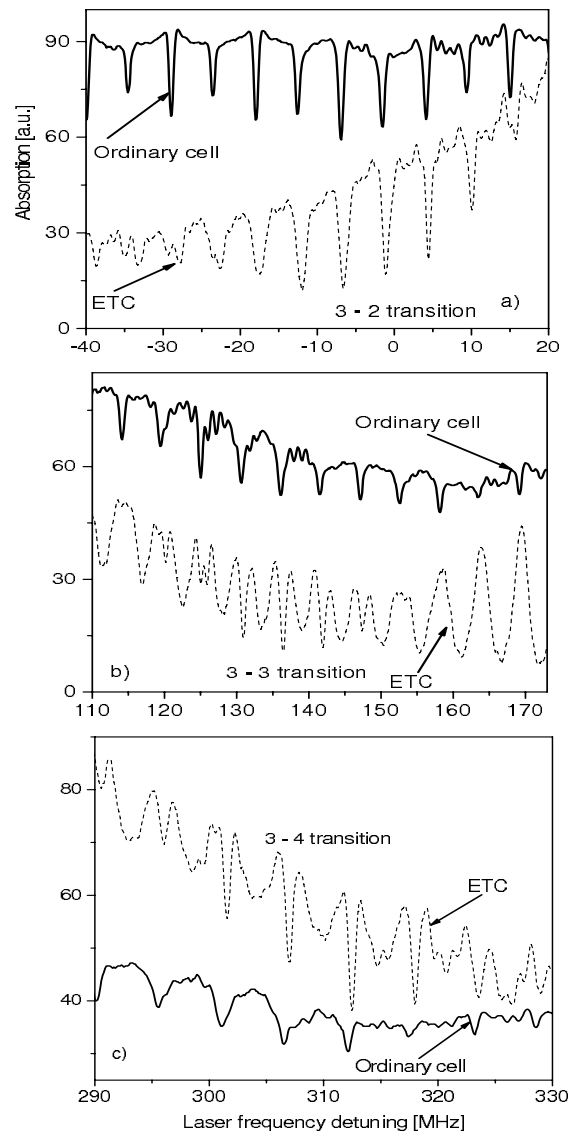}
\caption{Magneto-optical resonances superimposed on the ETC
absorption (dashed line) obtained by slow laser frequency detuning
in small intervals around the centers of $F_{g} = 3 \rightarrow
F_{e}$ = 2 (a), $F_{g} = 3 \rightarrow F_{e} = 3$ (b) and $F_{g} =
3 \rightarrow F_{e} = 4$ (c) transitions. A magnetic field is
applied and scanned around $B$ = 0 with a frequency two orders of
magnitude higher than that of the laser frequency scan. For
comparison of the $B$ = 0 positions, a simultaneous recording of
the resonances in ordinary cell is shown (solid line). Cs source
temperature is 120$^o$C, ETC thickness L = $\lambda$ and laser
power - 66mW/cm$^2$. Cs atoms are irradiated by linearly polarized
laser light.}
 \label{Fig4}
\end{figure}

As shown in the previous section, an important advantage of using
the ETC is that it is possible to study the behaviour of the
individual hf transitions. In order to test this property, the
following experiment is performed related to the magneto-optical
resonances. The laser frequency is slowly scanned over the hf
transition profile. During this slow scan, the magnetic field is
scanned around $B$ = 0 with a frequency about two orders of
magnitude higher than the frequency of laser scan. Such an
experiment allows the observation of a sequence of magneto-optical
resonances superimposed on the absorption profile of the
transition registered during the laser frequency detuning. The
minimum absorption of the dark resonances and the maximum
absorption of the bright resonances are observed at $B$ = 0.
First, the magneto-optical resonances at the three hf transitions
starting from the  $F_{g}$ = 3 level are examined
(Fig.\ref{Fig4}). In order to determine precisely the $B$ = 0
points, magneto-optical resonances are simultaneously registered
in the ordinary cell used for SA resonance observation
(Fig.\ref{Fig4}, solid lines). The ordinary cell is situated
outside the Helmholtz coils, so the magnetic field from the
Helmholtz coils there is not homogeneous and its value is lower
than that applied to the ETC. Hence, it is not possible to compare
the widths of magneto-optical resonances observed in both cells
but the determination of the $B$ = 0 points is correct.

Fig.\ref{Fig4}a illustrates the fact that magneto-optical dark
resonances are observed in a region of several tens of MHz around
the $F_{g} = 3 \rightarrow F_{e}$ = 2 transition center. This
interval is significantly narrower than the Doppler width of the
corresponding hf transition. The sign of the magneto-optical
resonance observed in the ETC corresponds to that observed in an
ordinary cell. A dark resonance is observed at the $F_{g}$ = 3
$F_{e}$ = 3 transition (Fig.\ref{Fig4}b), and it also exists in a
narrow region around the hf transition center. For this
transition, the resonance sign is also in agreement with the
observations in ordinary cells. From Fig.\ref{Fig4}b it can be
seen that the dark resonance observed for slow atoms is
superimposed on some other feature. The full profile of this
feature can be clearly distinguished for the atoms faster than
those contributing to the absorption around the hf transition
center. This type of Cs absorption dependence on the magnetic
field, which is broader than the dark resonance, has a maximum at
$B$ = 0 and is not related to coherent superposition of atomic
levels, will be discussed in more detail later in this Section.

As mentioned in the introduction, bright resonances are observed
at $F_{g} \rightarrow F_{e} = F_{g} + 1$ transitions of dilute
alkali atoms contained in the ordinary cells, where the collisions
between atoms can be neglected. Due to this, in the ETC, a narrow
bright resonance was expected for the $F_{g} = 3  \rightarrow
F_{e} = 4$ transition. However, at the intrinsically "bright"
$F_{g} = 3 \rightarrow F_{e} = 4$ transition, a dark resonance is
observed in the ETC. It should be pointed out that the $F_{g} = 3
\rightarrow F_{e} = 4 $ transition is an open one with a
transition probability significantly less than that of the closed
$F_{g} = 3 \rightarrow F_{e} = 2$ transition. As a result, in the
ordinary cell it was not possible to observe magneto-optical
resonances determined by this transition.

\begin{figure}[h]
\centering
\includegraphics[height=12cm, width=8cm, scale=0.5]{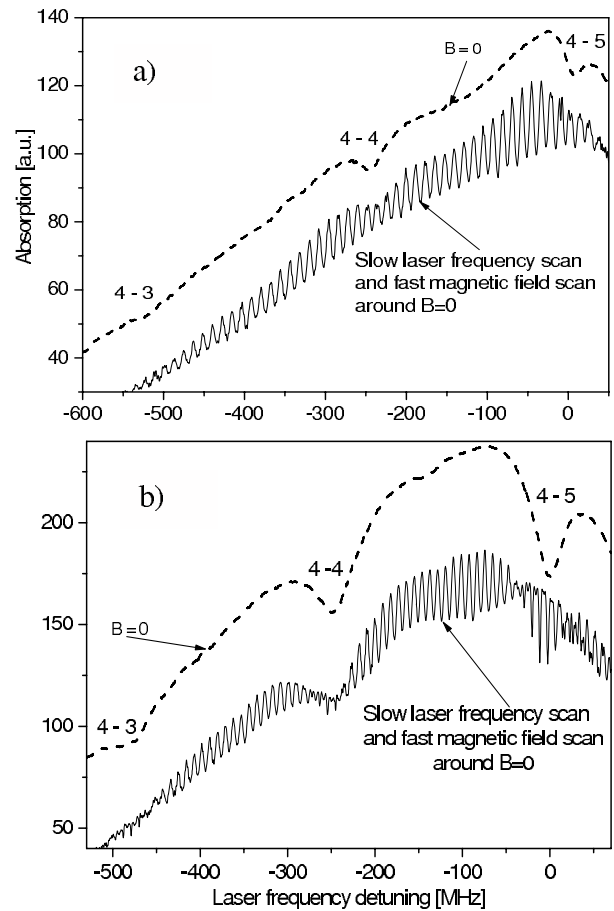}
\caption{Magneto-optical resonances superimposed on the ETC
absorption spectrum (solid line) and the corresponding part of the
ETC absorption spectrum (dashed line) at the $F_{g}$ = 4
absorption line for two different laser powers W=13mW/cm$^{2}$ (a)
and W=133mW/cm$^{2}$ (b). Cs source temperature is 131$^{o}$C, L
=$\lambda$, linearly polarized laser light.}
 \label{Fig5}
\end{figure}

In order to obtain more information about the behaviour of the
$F_{g} \rightarrow F_{e} = F_{g} + 1$ type of transitions in the
ETC, the $F_{g}$ = 4 set of transitions was investigated where the
$F_{g} = 4 \rightarrow  F_{e} = 5$ transition is closed and is the
strongest one. The obtained results are presented in
Fig.\ref{Fig5} where only the ETC signal is shown (in the presence
and in the absence of the magnetic field). In the first case
(Fig.\ref{Fig5}, solid line), the $B$ = 0 points are not noted and
we will specify the magneto-optical resonance sign. At low light
power (Fig.\ref{Fig5}a), the above mentioned broader features
predominate over the whole region of the presented absorption
spectrum, including the central regions of the three hf
transitions . At the centers of the hf transitions, only very
small-amplitude dark magneto-optical resonances are observed
superimposed on the tops of the profiles having maxima at $B$ = 0.
When increasing the light power (Fig.\ref{Fig4}b), the dark
resonance amplitude increases and, as in the case of the $F_{g}$ =
3 set of transitions, they are observed in a narrow interval
around the center of the hf transitions. More specifically, the
magneto-optical resonances are observed only within the frequency
intervals where the reduced absorption dips occur in the Cs
absorption spectrum at L = $\lambda$ (Fig.\ref{Fig5}, dashed curve
obtained at $B$ = 0). In the remaining regions of the light
frequency scan, the above mentioned features, which are assumed to
be of non-coherent origin are observed. Our experiment has shown
that for very low laser power, these features are observed over
the entire absorption spectra of both $F_{g}$ = 3, 4 sets of hf
transitions. Hence, the discussed features appear only under
conditions sufficient for the realization of single-photon
absorption. As the formation of the magneto-optical resonance
requires alignment/orientation of atoms (including several
absorption-fluorescence cycles), which is a much longer process,
we believe that the  formation of the non-coherent origin feature
is related to the large magnetic field scan (-90 G, +90 G) needed
to register the magneto-optical resonance. To understand this
effect, suppose that the laser light is in resonance with a
certain velocity class of atoms. As the width of the irradiating
light is small (3 MHz), only at $B$ = 0 all transitions starting
from different Zeeman sublevels will be in exact resonance with
the laser light. Taking into consideration the Zeeman splitting of
the ground state (0.35 MHz/G) and the excited state (0.93 MHz/G
for $F_{e}$ = 2, 0.00 for $F_{e}$ = 3, 0.37 MHz/G for $F_{e}$ = 4,
0.56 MHz/G for $F_{e}$ = 5) levels, it can be estimated that the
magnetic field increase will result in a significant shift from
the laser light frequency of the centers of optical transitions
that start from different ground-state Zeeman sublevels. Those
shifts will be proportional to the $m_{F}$ value, and only the
$m_{F_{g}} = 0 \rightarrow m_{F_{e}}$ = 0 transitions will stay in
resonance with the laser light during the magnetic field scan.

\begin{figure}[h]
\centering
\includegraphics[height=4cm, width=8cm, scale=0.5]{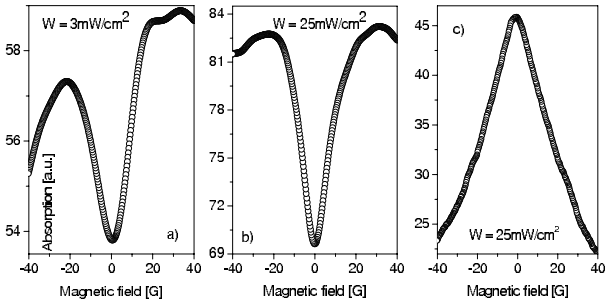}
\caption{ Illustration of dark magneto-optical resonances observed
at $F_{g} = 3 \rightarrow F_{e}$ = 2 (a) and $F_{g} = 3
\rightarrow F_{e} = 4$ (b) transitions for slow atoms, and
non-coherent feature (c) observed for fast atoms. Cs source
temperature is 112$^{o}$C and L = $\lambda$.}
 \label{Fig6}
\end{figure}

In general, the shifts of centers of different transitions are
significant compared to their natural width and might be
considered as a reason why the Cs atom absorption is reduced as
the magnetic field is detuned from zero-value, thus forming
features in the absorption dependence on magnetic field with
maxima at $B$ = 0. In this way, we attribute the discussed
features to the non-coherent dependence of absorption on the
magnetic field, whose origin are the shifts of the Zeeman
transition centers from the light frequency with the magnetic
field scan. Hence, only slow atoms, which suffer optical pumping
and saturation, are also involved in the formation of the
magneto-optical resonances. Fast atoms, which exhibit  much
shorter interaction time with the light, are mainly responsible
for the appearance of the non-coherent feature .

Fig.\ref{Fig6} illustrates the dependance of the magneto-optical
resonance and the non-coherent feature on magnetic field. The dark
magneto-optical resonance that is observed for slow atoms at all
$F_{g}  \rightarrow F_{e} = F_{g} - 1$, $F_{g}$ transitions is
illustrated in Fig.\ref{Fig6}a, for the closed $F_{g} = 3
\rightarrow F_{e} = 2$ transition. In Fig.\ref{Fig6}b, the dark
magneto-optical resonance that is observed in case of slow atoms
at the  $F_{g} = 3 \rightarrow F_{e} = 4$ transition is shown. The
case of fast atoms is illustrated in Fig.\ref{Fig6}c. Similar
results are observed by irradiating Cs atoms with circularly
polarized light and applying a magnetic field orthogonal to the
laser beam .

Coming back to Fig.\ref{Fig5}b, it should be particularly stressed
that for the closed $F_{g} = 4 \rightarrow F_{e} = 5$ transition,
also a dark resonance is evidenced in the case of the ETC. This
result is very interesting,  because precisely for this transition
a very well resolved bright resonance is observed in the ordinary
cell, which contains dilute Cs atoms \cite{And02}. However, it
turns out that the bright resonance is very sensitive to the
buffering of the  ordinary cell. It has been found \cite{And02}
and very recently confirmed \cite{Bra05,Bra06} that if the cell
containing alkali atoms is buffered by some noble gas, the
magneto-optical resonance at the $F_{g} = 4 \rightarrow F_{e} = 5$
transition is changed from a bright to a dark one. The theoretical
modeling has shown that this transformation of the resonance sign
can be attributed to the depolarization of the $F_{e}$ level by
collisions of alkali atoms with the buffer gas atoms. At the same
time, these collisions do not lead to depolarization of the
$F_{g}$ level, thus preserving the coherent superposition of the
ground-state magnetic sublevels introduced by the light at $B$ =
0. Note that the ordinary cell buffering does not reverse the sign
of the dark magneto-optical resonances observed on the $F_{g}
\rightarrow F_{e} = F_{g} - 1$, $F_{g}$ type of transitions.

\begin{figure}[tbp]
\includegraphics[width=8cm]{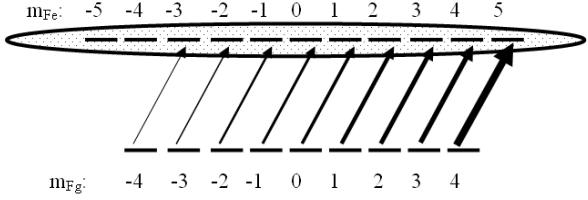}
\caption{  Illustration of the transformation of bright to dark
resonance at the $F_{g} = 4 \rightarrow  F_{e} = 5$ transition, in
case of depolarizing $F_{e}$ = 5 level collisions between Cs and
buffer gas atoms. } \label{Fig7}
\end{figure}

To easily picture the physical processes that lead to the
magneto-optical resonance sign reversal due to collisions between
alkali and buffer gas atoms in the ordinary cell, it is better to
consider Cs atoms irradiated by circularly polarized light. In
this case, the magneto-optical resonance is observed in absorption
(fluorescence) as a function of the magnetic field orthogonal to
the laser beam and scanned around $B$ = 0. Let us consider the
$F_{g} = 4 \rightarrow  F_{e} = 5$ transition on the D$_{2}$ line
of Cs (Fig.\ref{Fig7}). In the absence of depolarizing collisions
and in the presence of circularly polarized light ($\sigma^{+}$),
because the Clebsch Gordon coefficients increase for transitions
starting from $m_{F_{g}} = - 4$ to $m_{F_{g}} = 4$, most atoms
will circulate on the $m_{F_{g}} = 4 \rightarrow m_{F_{e}} = 5$
transition, which is the transition with the highest probability
of absorption for $\sigma^{+}$ polarization. Therefore, at $B$ = 0
a maximum in the absorption (fluorescence) will be observed. In
the presence of the magnetic field perpendicular to the atomic
orientation, part of the population of the $m_{F_{g}}$ = 4
sublevel will be redistributed to the other sublevels and the
absorption (fluorescence) will be decreased, which results in the
observation of a narrow resonance of enhanced absorption
(fluorescence) determined as a bright magneto-optical resonance.
However, if a buffer gas is added, a large fraction of the atoms
accumulated on the $m_{F_{e}}$ = 5 sublevel will be redistributed
among the other Zeeman sublevels of the $F_{e}$ = 5 level because
of  the depolarizing collisions between Cs and buffer gas atoms.
Taking into account the difference in the probabilities of the
transitions between the different Zeeman sublevels, it has been
shown \cite{And02} that this redistribution leads to an
accumulation (at $B$ = 0) of many atoms on the $m_{F_{g}}$ = - 4
sublevel, having the lowest probability for $\sigma^{+}$
excitation. Thus, the bright resonance observed in pure and dilute
Cs vapour transforms into a dark one when buffer gas is added to
the ordinary cell.

In analyzing the results related to the bright resonance sign
reversal due to the depolarization of the excited state, an
assumption has been made that in the ETC a similar depolarization
of the excited state can occur due to the long-range interaction
between alkali atoms and the two window surfaces of the ETC. In
support of such an assumption, note that during their interaction
with the light, atoms responsible for the magneto-optical
resonance formation fly a "long way" (on the order of a
millimeter) along the ETC windows, which are separated only by a
tiny gap very close to 852 nm. As an additional support to this
assumption, we report here the results of our measurement of the
polarization of the fluorescence collected in a direction
orthogonal to the laser beam. While in ordinary cell containing
pure Cs we measured a degree of fluorescence polarization $P$ =
0.24, in the ETC significant depolarization of the fluorescence
was observed yielding $P$ = 0.02. A theoretical model has been
developed to analyze the experimental observations that will be
presented in the following Section.

\section{Theoretical model}

\subsection{Introduction and definitions}

We consider the dipole interaction of an atom with a laser field in
presence of an external static magnetic field $\mathbf{B}$. We
assume that the atomic center of mass moves classically, which means
that the only effect of the dipole interaction of the atom with the
laser field is an excitation of a classically moving atom at the
internal transitions. In this case the internal atomic dynamics is
described by the semiclassical atomic density matrix $\rho $, which
parametrically depends on the classical coordinates of the atomic
center of mass.

We are interested in the optical Bloch equations (OBE) for the
density matrix elements $\rho _{g_{i}g_{j}},\rho _{g_{i}e_{j}},\rho
_{e_{i}g_{j}},\rho _{e_{i}e_{j}}$, where $g_{i}$ and $g_{j}$ label
atomic ground-state magnetic sublevels, and $e_{i}$ and $e_{j}$
excited-state magnetic sublevels. In writing the OBEs (see, for
example, \cite{Ste05}),
\begin{equation}
\imath \hbar \dfrac{\partial \rho }{\partial t} = \left[ \widehat{H},%
\rho \right] +\imath \hbar \widehat{R}\rho ,  \label{1}
\end{equation}%
we consider the relaxation $\widehat{R}$ due to spontaneous
emission, transit relaxation and due to interaction of atoms with
the cell walls -- both in non-elastic and elastic collisions. We
also assume, that different velocity groups do not mix in atom --
atom as well as atom -- wall interactions, since the atomic density
is sufficiently low.

The Hamiltonian
\begin{equation}
\widehat{H} = \widehat{H_{0}} + \widehat{H_{B}} + \widehat{V}
\label{1a}
\end{equation}%
includes the unperturbed atomic Hamiltonian $\widehat{H_{0}}$, which
depends on the internal atomic coordinates:
$\widehat{H_{0}}\left\vert \Psi _{n}\right\rangle =E_{n}\left\vert
\Psi _{n}\right\rangle $, $\widehat{H_{B}}$, the interaction of an
atom with the external magnetic field $\mathbf{B}$,  and the dipole
interaction operator $\widehat{V} = - \widehat{\mathbf{d}}\cdot \mathbf{E}%
\left( t\right) $, where $\widehat{\mathbf{d}}$ is the electric
dipole operator. The exciting light is described classically by a
fluctuating electric field $\varepsilon \left( t\right)$ of definite polarization $\mathbf{e}$:%
\begin{equation}
\mathbf{E}\left( t\right) =\varepsilon \left( t\right) \mathbf{e}%
+\varepsilon ^{\ast }\left( t\right) \mathbf{e}^{\ast }  \label{2a}
\end{equation}%
\begin{equation}
\varepsilon (t)=\left\vert \varepsilon _{\overline{\omega
}}\right\vert
e^{-\imath \Phi \left( t\right) -\imath \left( \overline{\omega }-\mathbf{k}%
_{\overline{\omega }}\mathbf{v}\right) t}  \label{2b}
\end{equation}%
with the center frequency of the spectrum $\overline{\omega }$\ and
the fluctuating phase $\Phi \left( t\right) $, which gives the
spectrum a finite bandwidth $\Delta \omega $. The lineshape of the
exciting light is
Lorentzian with FWHM $\triangle \omega $.

\subsection{Optical Bloch equations}

Writing OBEs explicitly for the density matrix element $\rho _{ij}$, we get:%
\begin{widetext}
\begin{eqnarray}
\dfrac{\partial \rho _{ij}}{\partial t} &=&-\dfrac{\imath }{\hbar
}\left[
\widehat{H},\rho _{ij}\right] +\widehat{R}\rho _{ij}=-\dfrac{\imath }{\hbar }%
\left[ \widehat{H_{0}},\rho _{ij}\right] +\dfrac{\imath }{\hbar
}\left[
\widehat{\mathbf{d}}\cdot \mathbf{E}\left( t\right) ,\rho _{ij}\right] +%
\widehat{R}\rho _{ij}  =\notag \\
&=&-\imath \omega _{ij}\rho _{ij}+\dfrac{\imath }{\hbar
}\mathbf{E}\left( t\right) \underset{k}{\sum }\left(
\mathbf{d}_{ik}\cdot \rho _{kj}-\rho
_{ik}\cdot \mathbf{d}_{kj}\right) +\widehat{R}\rho _{ij}  =\notag \\
&=&-\imath \omega _{ij}\rho _{ij}+\dfrac{\imath }{\hbar }\varepsilon
\left(
t\right) \underset{k}{\sum }d_{ik}\rho _{kj}+\dfrac{\imath }{\hbar }%
\varepsilon ^{\ast }\left( t\right) \underset{k}{\sum }d_{ik}^{\ast }\rho _{kj}-%
\dfrac{\imath }{\hbar }\varepsilon \left( t\right) \underset{k}{\sum }%
d_{kj}\rho _{ik}-\dfrac{\imath }{\hbar }\varepsilon ^{\ast }\left(
t\right) \underset{k}{\sum }d_{kj}^{\ast }\rho _{ik}+\widehat{R}\rho
_{ij} \label{3}
\end{eqnarray}
\end{widetext}%
where $\omega _{ij}=\dfrac{E_{i} - E_{j}}{\hbar }$, and the
transition dipole
matrix elements ${d}_{ij} = \left\langle i\left\vert \mathbf{d}%
\right\vert j\right\rangle $ and $d_{ij} = \left\langle i\left\vert \mathbf{d}\cdot \mathbf{e} %
\right\vert j\right\rangle $ can be calculated using the standard
angular momentum algebra \cite{Auz05,Var88,Zar88}.

\subsection{Simplification of the OBEs}

Next, we apply the following procedure, which simplifies OBE. First,
we use the rotating wave approximation \cite{All75}. After that we
have the equations, which are stochastic differential equations
\cite{Kam76} with stochastic variable $\dfrac{\partial \Phi \left(
t\right) }{\partial t}$. In experiment we measure time averaged
(stationary) values. Therefore, we need to perform the statistical
averaging of the above equations. In order to do that, we solve the
equations for optical coherences and then take a formal statistical
average over the fluctuating phases. Finally we apply the
"decorrelation approximation" (which in general is valid only for
Wiener-Levy-type phase fluctuations) and assume a "phase diffusion"
model for the description of the dynamics of the fluctuating phase.
Thus we obtain a phase averaged OBE. For a detailed description of
the procedure of statistical averaging, decorrelation approximation,
Wiener-Levy-type phase fluctuations, and phase-diffusion model, see
\cite{Blu04}\ and references cited therein.

\subsection{Steady-state excitation}

Now we assume, that atoms in the cell reach the stacionary
excitation conditions (steady-state). Under such conditions all
density matrix elements in the OBE become time independent. In
these conditions we can eliminate the optical coherences $\rho
_{g_{i}e_{j}}$ and $\rho _{e_{i}g_{j}}$ \cite{Bri99,Ste05} from
OBE.
Thus, for optical coherences we obtain the following explicit expressions:%
\begin{equation}
\widetilde{\rho _{g_{i}e_{j}}}=\dfrac{\imath }{\hbar
}\dfrac{\left\vert \varepsilon _{\overline{\omega }}\right\vert
}{\Gamma _{R}+\imath \Delta _{e_{j}g_{i}}}\left(
\underset{e_{k}}{\sum }d_{g_{i}e_{k}}^{\ast }\rho
_{e_{k}e_{j}}-\underset{g_{k}}{\sum }d_{g_{k}e_{j}}^{\ast }\rho
_{g_{i}g_{k}}\right)  \label{4a}
\end{equation}%
\begin{equation}
\widetilde{\rho _{e_{i}g_{j}}}=\dfrac{\imath }{\hbar
}\dfrac{\left\vert \varepsilon _{\overline{\omega }}\right\vert
}{\Gamma _{R}-\imath \Delta
_{e_{i}g_{j}}}\left( \underset{g_{k}}{\sum }d_{e_{i}g_{k}}\rho _{g_{k}g_{j}}-%
\underset{e_{k}}{\sum }d_{e_{k}g_{j}}\rho _{e_{i}e_{k}}\right),
\label{4b}
\end{equation}%
where
\begin{equation}
\Delta _{ij} = \overline{\omega } - \mathbf{k_{\overline{\omega
}}}\mathbf{v} - \omega _{ij}  \label{5}
\end{equation}%
and $\mathbf{k_{\overline{\omega }}}\mathbf{v}$ represents the
Doppler shift of the transition energy of an atom due to its spatial
motion. Here $\mathbf{k}$ is the wave vector of the excitation light
and $\mathbf{v}$ is the atom velocity. In the present study it was
assumed that in the ETC $\mathbf{k_{\overline{\omega }}}\mathbf{v} =
0$. $\Gamma _{R}$ describes effective relaxation,%
\begin{equation}
\Gamma _{R} = \frac{\Gamma }{2} + \frac{\Delta \omega }{2} + \gamma
+ \Gamma _{col},  \label{6}
\end{equation}%
where $\Gamma $ describes spontaneous relaxation from level $e$,
$\gamma $ describes the relaxation rate, $\Delta \omega $, the
relaxation due to finite linewidth of the laser (this relaxation is
the consequence of the statistical averaging over the laser field
fluctuations), and $\Gamma _{col}$ describes the relaxation due to
elastic atomic collisions with the cell walls.

\subsection{Rate equations for Zeeman coherences}

By substituting the expressions for the optical coherences Eqs.
(\ref{4a}, \ref{4b}) in the equations for the Zeeman coherences, we
arrive at the rate equations for Zeeman coherences only (see
\cite{Blu04}):
\begin{widetext}
\begin{center}
\begin{eqnarray}
\dfrac{\partial \rho _{g_{i}g_{j}}}{\partial t} &=&0=-\imath \omega
_{gigj}\rho _{g_{i}g_{j}}-\gamma \rho _{g_{i}g_{j}}+\underset{e_{i}e_{j}}{%
\sum }\Gamma _{g_{i}g_{j}}^{e_{i}e_{j}}\rho _{e_{i}e_{j}}+\lambda
\delta
\left( g_{i},g_{j}\right) + \notag \\
&& + \dfrac{\left\vert \varepsilon _{\overline{\omega }}\right\vert
^{2}}{\hbar ^{2}}\underset{e_{k},e_{m}}{\sum }\left(
\dfrac{1}{\Gamma _{R}+\imath \Delta _{e_{m}g_{i}}}+\dfrac{1}{\Gamma
_{R}-\imath \Delta _{e_{k}g_{j}}}\right)
d_{g_{i}e_{k}}^{\ast }d_{e_{m}g_{j}}\rho _{e_{k}e_{m}} - \notag \\
&&-\dfrac{\left\vert \varepsilon _{\overline{\omega }}\right\vert ^{2}}{%
\hbar ^{2}}\underset{e_{k},g_{m}}{\sum }\left( \dfrac{1}{\Gamma
_{R}-\imath
\Delta _{e_{k}g_{j}}}d_{g_{i}e_{k}}^{\ast }d_{e_{k}g_{m}}\rho _{g_{m}g_{j}}+%
\dfrac{1}{\Gamma _{R}+\imath \Delta
_{e_{k}g_{i}}}d_{g_{m}e_{k}}^{\ast }d_{e_{k}g_{j}}\rho
_{g_{i}g_{m}}\right)  \label{7a}
\end{eqnarray}

\begin{eqnarray}
\dfrac{\partial \rho _{e_{i}e_{j}}}{\partial t} &=&0=-\imath \omega
_{e_{i}e_{j}}\rho _{e_{i}e_{j}}-\gamma \rho _{e_{i}e_{j}}-\Gamma
\rho
_{e_{i}e_{j}}-\Gamma _{col}\rho _{e_{i}e_{j}}  + \notag \\
&& + \Gamma _{col}\cdot \delta \left( e_{i},e_{j}\right) \cdot \underset{e_{k}}{%
\sum }\frac{N_{e_{k}e_{i}}}{\underset{e_{m}}{\sum }N_{e_{k}e_{m}}}\rho _{e_{k}e_{k}} + \notag \\
&& + \dfrac{\left\vert \varepsilon _{\overline{\omega }}\right\vert
^{2}}{\hbar ^{2}}\underset{g_{k},g_{m}}{\sum }\left( \dfrac{1}{
\Gamma _{R}-\imath \Delta _{e_{i}g_{m}} }+\dfrac{1}{ \Gamma
_{R}+\imath \Delta _{e_{j}g_{k}} }\right)
d_{e_{i}g_{k}}d_{g_{m}e_{j}}^{\ast }\rho
_{g_{k}g_{m}} - \notag \\
&&-\dfrac{\left\vert \varepsilon _{\overline{\omega }}\right\vert ^{2}}{%
\hbar ^{2}}\underset{g_{k},e_{m}}{\sum }\left( \dfrac{1}{\Gamma
_{R}+\imath
\Delta _{e_{j}g_{k}}}d_{e_{i}g_{k}}d_{g_{k}e_{m}}^{\ast }\rho _{e_{m}e_{j}}+%
\dfrac{1}{\Gamma _{R}-\imath \Delta _{e_{i}g_{k}}}%
d_{e_{m}g_{k}}d_{g_{k}e_{j}}^{\ast }\rho _{e_{i}e_{m}}\right)
\label{7b}
\end{eqnarray}
\end{center}
\end{widetext}
Here $\Gamma _{g_{i}g_{j}}^{e_{i}e_{j}}$ describes the spontaneous
relaxation from $\rho _{e_{i}e_{j}}$ to $\rho _{g_{i}g_{j}}$,
$\lambda $ describes the rate at which "fresh" atoms move into
interaction region in the transit relaxation process, $\delta \left(
i, j\right) $ is the Dirac delta symbol, but $N_{ij}$ describes the
relative rate of the elastic interaction of atoms with the walls of the ETC:%
\begin{equation}
N_{ij} = \dfrac{\Delta _{col}^{2}}{\Delta _{col}^{2} + \left\vert
\omega _{ij}\right\vert ^{2}}  \label{8}
\end{equation}

\subsection{Model for elastic collisions}

When atoms move through the cell, they may experience elastic and
inelastic interactions with the walls of the ETC. As was shown in
\cite{Ham05}, the  ETC provides the possibility to explore the
long-range atom-surface van der Waals interaction and modification
of atomic dielectric resonant coupling under the influence of the
coupling between the two neighboring dielectric media, and even the
possible modification of interatomic collision processes under the
effect of confinement. Besides,  if an atom is "flying" very close
to the surface, it can even experience the periodic potential coming
from the crystalline surface of the ETC. We will speak about these
effect as the effect of elastic collisions occurring with the rate
$\Gamma _{col}$.

We assume the following model for elastic collisions. First, these
collisions do not affect ground state (Eq. \ref{7a}), as in the
ground state $L_{g} = 0$, and we
assume that these collisions cannot "turn" the spin -- neither electronic spin $S$%
, nor nuclear spin $I$. Thus, collisions affect excited-state Zeeman
coherences and populations (Eq. \ref{7b}), as well as optical
coherences Eqs. (\ref{4a}, \ref{4b}).

The second assumption is that elastic collisions redistribute
populations only among excited-state magnetic sublevels, see
 Eq. (\ref{7b}), which means that both the optical coherences and the
 excited-state Zeeman coherences are destroyed completely with the rate
$\Gamma _{col}$.

Finally we want to note that the external $\mathbf{B}$-field not
only causes magnetic sublevel splitting $\omega_{e_{i}e_{j}}$  and
$\omega_{g_{i}g_{j}}$, but also alters the dipole transition matrix
elements by mixing the hf levels with the same magnetic quantum
number $m$, but with different hf level angular momentum quantum
number $F$.
\begin{equation}
|e_{i}\rangle = \sum_{F_{e}} c_{i}^{(e)} |F_{e} m_{i}\rangle
\label{9a}
\end{equation}
\begin{equation}
|g_{i}\rangle = \sum_{F_{g}} c_{i}^{(g)} |F_{g} m_{i}\rangle
\label{9b}
\end{equation}

The numerical values of the magnetic sublevel splitting energies
$\omega_{e_{i}e_{j}}$  and $\omega_{g_{i}g_{j}}$, as well as the hf
state mixing coefficients $c_{i}^{(e)}$  and $c_{i}^{(g)}$  are
obtained by diagonalization of the magnetic field interaction
Hamiltonian $\widehat{H_{B}}$ Eq. (\ref{1a}).

The third assumption is the phenomenological model for the
redistribution of the excited-state population due to collisions,
namely Eq. (\ref{8}). The quantity $\sum_{e_{m}} N_{e_{k}e_{m}}$  in
Eq. (\ref{7b}) is the normalization coefficient. As can be seen from
Eq. (\ref{8}), we assume that the distribution of the probability
for the atomic interaction with the ETC walls to mix populations in
the excited state of the atoms has a Lorentzian shape with FWHM
$\Delta_{col}$ . Thus $\Delta_{col}$ could be called the effective
width of the elastic collisions.

\section{Comparison between theoretical and experimental results}
Based on the  model developed for the ETC, the magneto-optical
resonance profiles have been computed for laser frequency detuning
over all hf transitions and for both types of excitations, that
is, by linearly and circularly polarized light. In the
calculations, the following parameters related to the experimental
conditions have been used: $\Gamma  $ = 6 MHz, $\gamma$ = 5 MHz,
$\Delta \omega$  = 3 MHz, $\Gamma_{col}$ = 200 MHz, $\Delta_{col}$
= 50 MHz.

In all the cases considered, the theoretical results are at least in
qualitative agreement with the experimental observations. As in the
experiment, reduced absorption magneto-optical resonances are
predicted by the theory for all $F_{g} \rightarrow F_{e} = F_{g} -
1$, $F_{g}$ transitions.

\begin{figure*}[tbp]
\includegraphics[width=4.5cm]{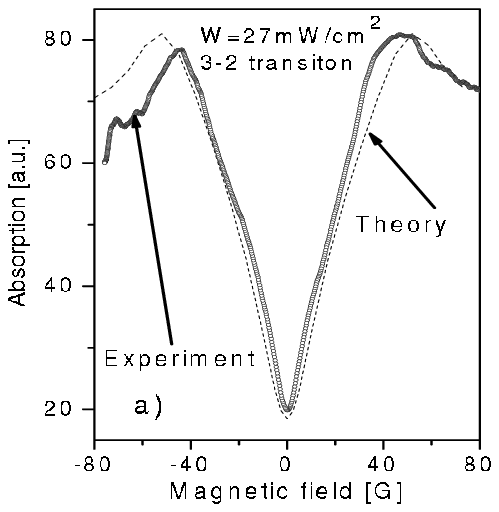}
\includegraphics[width=5.2cm]{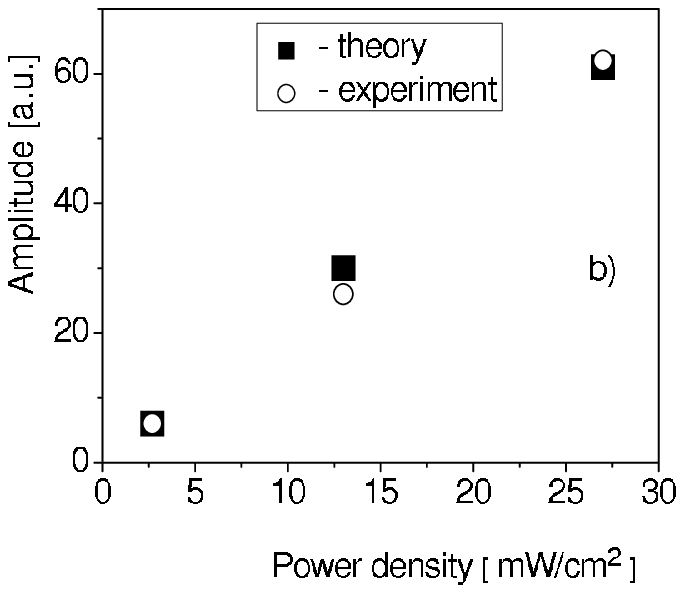}
\includegraphics[width=5.2cm]{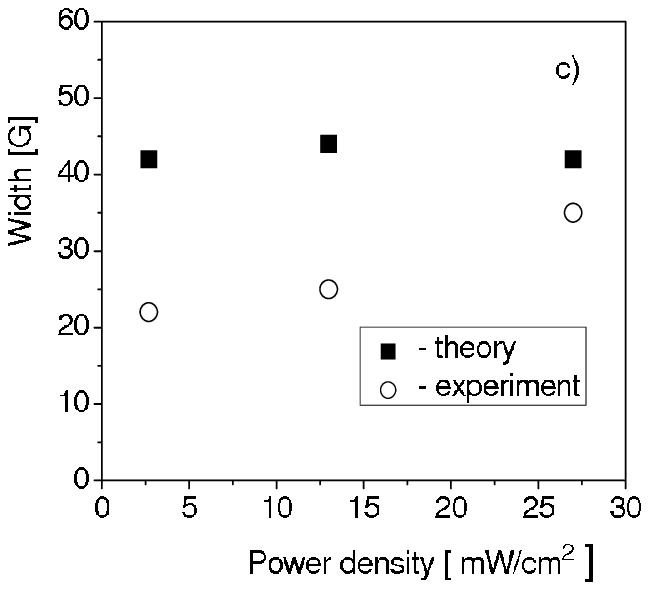}
\caption{  Comparison between theory and experiment, for $F_{g} =
3 \rightarrow F_{e} = 2$ transition and excitation by linearly
polarized light: (a) - magneto-optical resonance experimental and
theoretical profiles; (b) - magneto-optical resonance amplitude as
function of the light power density; (c) - magneto-optical
resonance width as a function of power density. } \label{Fig8}
\end{figure*}

The inclusion in the theoretical model of the influence of the
cell walls on the polarization of the excited atomic level has as
a result the sign reversal of the bright magneto-optical
resonance. Mixing of the excited state Zeeman sublevels according
to Eq. (\ref{7b}) and (\ref{8}) is assumed, which results in
atomic accumulation in the ground-state Zeeman sublevels with the
lowest probability of excitation (at $B$ = 0) for the $F_{g}
\rightarrow F_{e} = F_{g} + 1$ transitions. This atomic
accumulation in "the less favored" ground sublevel is the reason
for the bright magneto-optical resonance sign reversal. The
experimentally observed reduced-absorption magneto-optical
resonances at this type of transition supports our theoretical
result.  We would like to remind the reader, that in the case of
the $F_{g} \rightarrow  F_{e} = F_{g} + 1$ transitions, without
the cell window influence, atomic excitation by light of any
polarization would results in atomic accumulation in the
ground-state Zeeman sublevel with the highest probability of
excitation (at $B$ = 0), and hence in the observation of
enhanced-absorption, bright magneto-optical resonance.

 Let us discuss the behaviour of particular magneto-optical resonances when
the laser frequency is tuned to the central frequencies of the
different hf transitions. In an interval of a few tens of MHz
around the $F_{g} = 3 \rightarrow F_{e}$ = 2 transition, a dark
resonance is experimentally observed and is shown in
Fig.\ref{Fig8} together with the theoretically simulated profile
of the resonance. Here, Cs atoms are irradiated by linearly
polarized light and the magnetic field is orthogonal to the
polarization vector and to the light propagation direction. A good
quantitative agreement between the theoretical and the
experimental results can be seen. The resonance amplitude
increases with the light power density and here the agreement of
the theory with the experiment is very good. Some discrepancy
still remains between the theoretical and experimental resonance
widths for low laser power, which can be due to the residual
Doppler broadening not accounted for in the model. When the laser
power is increased, the transition saturation effects start to
dominate over the Doppler broadening. The $F_{g} = 3 \rightarrow
F_{e} = 3$ transition behaves similar to that of the $F_{g} = 3
\rightarrow  F_{e} = 2$ transition. Here also a dark
magneto-optical resonance is observed, and also the theoretical
profile parameters are in agreement with those of the experimental
profiles.

\begin{figure*}[tbp]
\includegraphics[width=4.4cm]{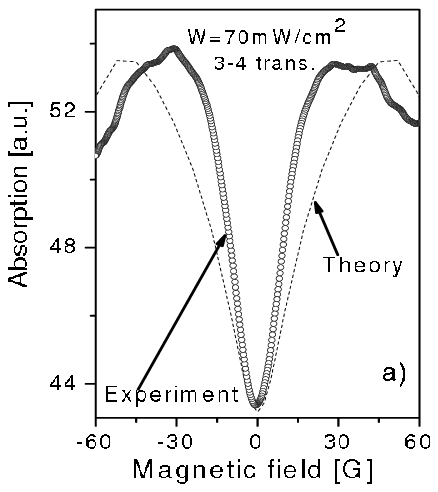}
\includegraphics[width=5.7cm]{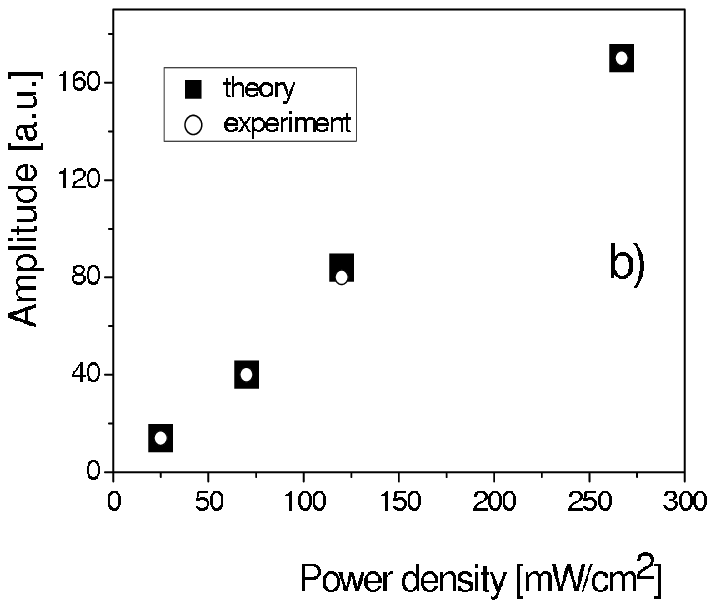}
\includegraphics[width=5.4cm]{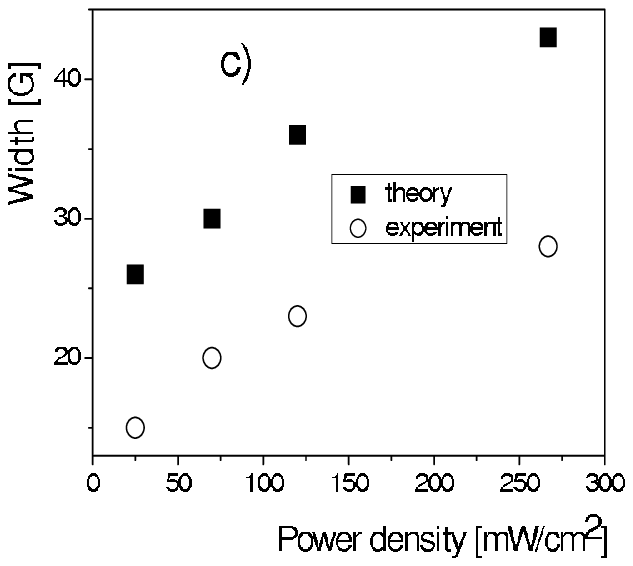}
\caption{  Comparison between theory and experiment, for $F_{g} =
3 \rightarrow F_{e} = 4$ transition and excitation by linearly
polarized light: (a) - magneto-optical resonance experimental and
theoretical profiles; (b) - magneto-optical resonance amplitude as
a function of the light power density; (c) - magneto-optical
resonance width as a function of power density.} \label{Fig9}
\end{figure*}

It should be pointed out that the magneto-optical resonance width in
the ETC is an order of magnitude larger than that obtained in the
ordinary cell. This broadening can be attributed to the larger value
of the transit relaxation rate due to the nanometric scale of the
ETC thickness as compared to the cm-dimensions of the ordinary cell.
Note that the ETC is not shielded against the laboratory magnetic
field because of technical reasons connected with its shape and
because the magnetic field must be scanned in quite a large
interval. But as the laboratory magnetic field is less than a Gauss,
the lack of shielding should not influence significantly the
resonance width. Moreover, an experiment with an unshielded ordinary
cell situated very close to the ETC shows a magneto-optical
resonance with less than 1 G width, which proves that the laboratory
magnetic field cannot be the main reason for the resonance
broadening.

\begin{figure*}[tbp]
\includegraphics[width=3.2cm]{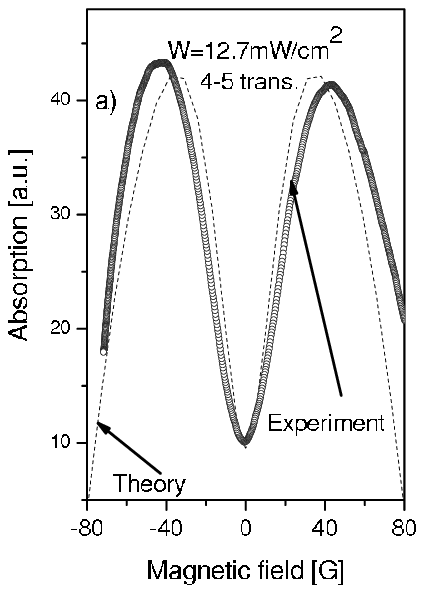}
\includegraphics[width=6.3cm]{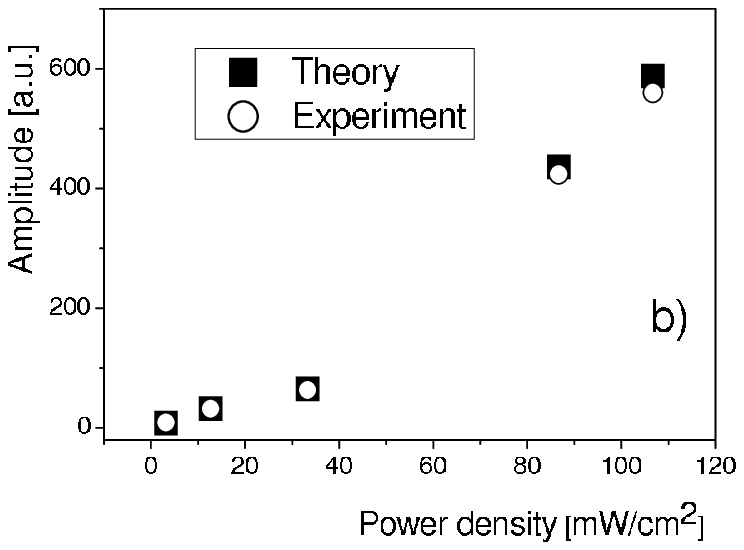}
\includegraphics[width=5.8cm]{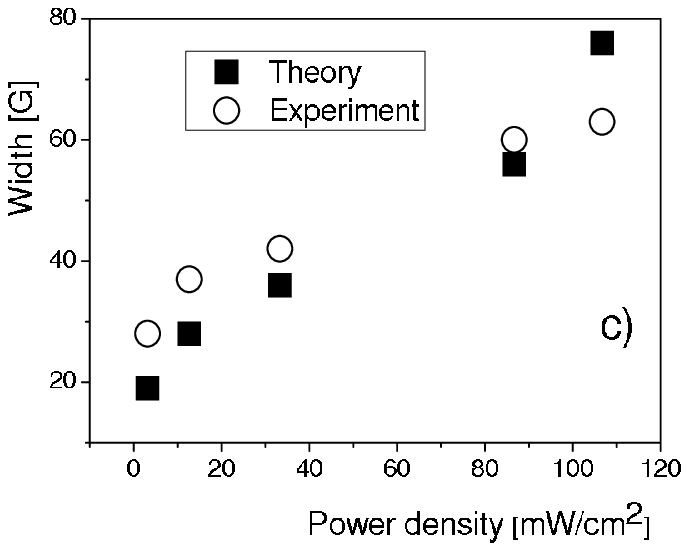}
\caption{Comparison between  theory and experiment, for $F_{g} = 4
\rightarrow F_{e} = 5$ transition and excitation by circularly
polarized light: (a) - magneto-optical resonance experimental and
theoretical profiles; (b) - magneto-optical resonance amplitude as
a function of the light power density; (c) - magneto-optical
resonance width as a function of power density.} \label{Fig10}
\end{figure*}

The $F_{g} = 3 \rightarrow  F_{e} = 4$ transition represents an
interesting case. As mentioned above for the case of the ordinary
cell, because of the strong overlapping of the Doppler profiles of
the hf transitions and the weakness of the $F_{g} = 3 \rightarrow
F_{e} = 4$ open transition, when tuning the laser frequency in
resonance with this transition, it has not been possible to
observe enhanced-absorption (bright) magneto-optical resonances.
In the ETC, however, a large amplitude magneto-optical resonance
is observed when the laser frequency is tuned to the center of the
transition. But instead of a bright resonance, a dark resonance
appears at the $F_{g} = 3 \rightarrow F_{e} = 4$ transition, in
agreement with the theoretical model. Both the theoretical and
experimental resonance profiles are illustrated in Fig.\ref{Fig9}
for irradiation by linearly polarized light. Here also the scanned
magnetic field is oriented in a direction orthogonal to the light
polarization and propagation direction. In the case of this
resonance, the theory is also in very good quantitative agreement
with the experiment related to the dependance of the resonance
amplitude on the laser power density. Here, we observe in the
magneto-optical resonance width a discrepancy between the theory
and the experiment that is even larger than in the previous case.
The experimentally observed resonance is significantly narrower
than the theoretical one. The origin of this discrepancy can be
attributed to the fact that the $F_{g} = 3 \rightarrow F_{e} = 4$
transition is an open transition. Therefore, a significant
population loss occurs from the $F_{g}$ = 3 level  to the $F_{g}$
= 4 level as a result of the hf optical pumping through
spontaneous transitions from the $F_{e}$ = 4 level. In
\cite{Ren99,Ren99b,Ren98}, it has been shown that, in the case of
open hf transitions, a significant coherent resonance narrowing
takes place, because the resonance destruction is more effective
at the wings, while at the center the resonance is more resistant
to the loss.

Based on the results presented in Fig.\ref{Fig9}, we can conclude
that, unlike the dark magneto-optical resonance, the bright
resonance is very sensitive, not only to the depolarization
collisions of alkali atoms with those of the buffer gas in the
ordinary cell, but also to the surface - atom interactions, about
which it can provide valuable information.

In order to confirm the sensitivity of the $F_{g} \rightarrow
F_{e} = F_{g} + 1$ type of transitions to the depolarizing
influence of the ETC windows, the behaviour of the closed and
high-probability $F_{g} = 4 \rightarrow F_{e} = 5$ transition is
investigated in detail by irradiating Cs atoms with linearly or
circularly polarized light. In the first case the scanned magnetic
field is oriented orthogonally to the atomic alignment, while in
the second case, orthogonally to the orientation of atoms. In both
cases reduced-absorption magneto-optical resonance is observed
experimentally for atoms confined in the ETC. Expanding the
discussion related to Fig.\ref{Fig7}, it can be pointed out that
in case of dilute atoms confined in the ordinary cell at $B$  = 0,
irradiation by linearly polarized light also results in atomic
accumulation in the ground-state Zeeman sublevels with the highest
excitation probabilities. And similar to the case of circular
polarization, also in case of linear polarization the excited
state depolarization by alkali atom collisions with buffer gas
atoms results in a sign reversal of the magneto-optical resonance.

The results obtained from studies with both polarization
excitations and with the laser frequency tuned to  the central
frequency of the $F_{g} = 4 \rightarrow  F_{e} = 5$ transition,
confirm the assumed analogy between the excited state
depolarization by atomic collisions (for the ordinary cell) and
its depolarization by the influence of the electrical potential of
the ETC window. In Fig.\ref{Fig10}, a comparison between the
experimental and theoretical results is presented for the $F_{g} =
4 \rightarrow  F_{e} = 5$ transition, in case of Cs atoms
irradiated by circularly polarized light with a magnetic field
applied in a direction orthogonal to the light propagation. It can
be seen that in relation to the magneto-optical resonance width,
the agreement between the theory and the experiment is better for
the $F_{g} = 4 \rightarrow F_{e} = 5$ transition. The reason for
this can be the fact that, unlike the $F_{g} = 3 \rightarrow F_{e}
= 4$ transition, the $F_{g} = 4 \rightarrow F_{e} = 5$ transition
is a closed one, and so the resonance narrowing due to the
population loss is not expected for the last transition.

\section{Conclusion}
We have presented an experimental and theoretical study of the
ground-state magneto-optical resonances prepared in Hanle
configuration on the D$_{2}$ line of Cs. In a cm-scale ordinary cell
containing Cs vapour, the hyperfine transitions starting from a
single ground-state level are strongly overlapped, which is a reason
for the mixing of the contribution of different hf transitions that
are responsible for the dark and bright magneto-optical resonances.
It is shown that the utilization of an Extremely Thin Cell with
thickness equal to the wavelength of the irradiating light allows
one to examine the formation of a magneto-optical resonance
 on the individual hf transitions due to the
experimentally proven fact that only very slow atoms possess enough
interaction time with the light to form the magneto-optical
resonance. The fast atoms, exhibiting mainly a single act of
absorption during their interaction with the light, do not
contribute to the magneto-optical resonance formation.

It is shown that in the ETC, dark (reduced absorption)
magneto-optical resonances are observed at $F_{g} \rightarrow  F_{e}
= F_{g} - 1$, $F_{g}$ transitions, much like in ordinary cells.
However in case of the $F_{g} \rightarrow F_{e} = F_{g}+ 1$
transitions, Cs atoms confined in ETC exhibit completely different
behaviour as compared to those contained in an ordinary cell. For
the latter, bright (enhanced absorption) magneto-optical resonances
have been observed in dilute Cs vapour in the ordinary cell. As a
result of our study we report on the bright resonance sign reversal
in Cs atoms confined in the ETC. Both ($F_{g} = 3 \rightarrow F_{e}
= 4$ and $F_{g} = 4 \rightarrow  F_{e} = 5$) intrinsically "bright"
hf transitions are responsible for dark magneto-optical resonance
formation in the ETC.

A theoretical model is proposed based on the optical Bloch equations
that involves the elastic interaction processes of atoms in the ETC
with its walls. The assumed elastic collisions of Cs atoms with the
cell walls do not affect the ground state of alkali, do not reorient
the electronic and nuclear spins of the atom, but do affect the
excited-state Zeeman coherences and populations, as well as the
optical coherences. The involvement of elastic collisions of this
type results in depolarization of the Cs excited state that had been
polarised by the exciting radiation. This depolarization leads to
the accumulation of atomic population in ground-state Zeeman
sublevels with the lowest probability of excitation, which is
opposite to the situation in the ordinary cell, where atoms
accumulate on the Zeeman sublevel possessing the largest probability
of excitation. Hence, the  influence of the ETC wall on atomic
polarization leads to the sign reversal of bright resonance. Using
the proposed model, the magneto-optical resonance amplitude and
width dependences on laser power are calculated and compared with
the experimental ones. The numerical results are in good agreement
with the experiment.

The obtained results show that the magneto-optical resonances
observed in the ETC could potentially be applied to study
atom-surface interactions.

\section{Acknowledgements}

The authors are grateful to Prof. D. Sarkisyan for providing the
ETC and for the number of extremely useful discussions, as well as
we acknowledge  partial support from the INTAS programm (grant: 06
- 1000017 - 9001). C. A., S. C., L. P. and D. S. acknowledge
support from the Bulgarian Fund for Scientific Research (grant
F-1404/04). A. A., M. A. and K. B. acknowledge support from EU FP6
TOK project LAMOL and European Regional Development Fund project
Nr. 2.5.1./000035/018. A. A. and K. B gratefully acknowledge
support from the European Social Fund. S.C. appreciates very
helpful discussions with Prof. D. Bloch, Prof. M. Ducloy and Prof.
Yu. Malakyan  related to the particular properties of the atoms
confined in the ETC and magneto-optical resonances.

\bibliography{apssamp}

\end{document}